\documentclass[andp2012,twocolumn]{revtex4}
\usepackage{dcolumn}
\usepackage{bm}
\usepackage{graphicx}
\usepackage{amsmath}
\usepackage{latexsym}
\usepackage{amsfonts}
\usepackage{amssymb}
\usepackage{array}
\usepackage{epsfig}
\usepackage{times}
\usepackage{txfonts}
\usepackage{epstopdf}
\usepackage{pifont}
\usepackage{dsfont}
\usepackage{amscd}
\usepackage{amsfonts}
\usepackage{textcomp}

\newcommand{\be}{\begin{equation}}
\newcommand{\ee}{\end{equation}}
\newcommand{\bea}{\begin{eqnarray}}
\newcommand{\eea}{\end{eqnarray}}

\begin{document}
\title{Enhancing squeezing and nonclassicality of light in atom-optomechanical systems}
\author{Yanqiang Guo$^{1,2}$, Xiaomin Guo$^{1,2}$, Pu Li$^{1,2}$, Heng Shen$^4$, Jing Zhang$^3$, and Tiancai Zhang$^3$}
\affiliation{$^1$Key Laboratory of Advanced Transducers and Intelligent Control System, Ministry of Education, Taiyuan University of Technology, Taiyuan 030024, China\\
$^2$Institute of Optoelectronic Engineering, College of Physics and Optoelectronics, Taiyuan University of Technology, Taiyuan 030024, China\\
$^3$State Key Laboratory of Quantum Optics and Quantum Optics Devices, Institute of Opto-Electronics, Shanxi University, Taiyuan 030006, China \\
$^4$Clarendon Laboratory, University of Oxford, Parks Road, Oxford OX1 3PU, United Kingdom}

\begin{abstract}
Quadrature squeezing of light is investigated in a hybrid atom-optomechanical system comprising a cloud of two-level atoms and a movable mirror mediated by a single-mode cavity field. When the system is at high temperatures with quadrature fluctuations of light much above the standard quantum limit (SQL), excitation counting on the collective atomic state can effectively reduce the light noise close to the SQL. When the system is at low temperatures, considerable squeezing of light below the SQL is found at steady state. The squeezing is enhanced by simply increasing the atom-light coupling strength with the laser power optimized close to the unstable regime, and further noise reduction is achieved by decreasing various losses in the system. The presence of atoms and excitation counting on the atoms lessen the limitation of thermal noise, and the squeezing can be achieved at environment temperature of the order K. The nonclassicality of the light, embodied by the negative distributions of the Wigner function, is also studied by making non-Gaussian measurements on the atoms. It is shown that with feasible parameters excitation counting on the atoms is effective in inducing strongly optical nonclassicality.

\end{abstract}
\date{\today}
\maketitle

Squeezed light, with quantum fluctuations below the level of vacuum noise or the standard quantum limit (SQL), is particularly useful for ultra-sensitive force measurements~\cite{caves}, gravitational wave detection~\cite{LIGO}, and is also an important resource in quantum information science for continuous-variable information processing~\cite{loock}. Squeezed light was first generated by using atomic sodium as a nonlinear medium~\cite{sodium}, and then followed by employing optical fibers~\cite{fiber} and nonlinear crystals~\cite{crystal}. Significant squeezing, up to $15$ dB, has been obtained in the laboratory~\cite{15db}.

Over twenty years ago, an optomechanical cavity was proposed to generate the so-called ponderomotive squeezing of light~\cite{Fabre,Mancini} due to its similarity to a Kerr nonlinear medium. The mechanical element is shifted by the radiation pressure, proportionally to the intracavity intensity of optical field, which in turn modulates the phase of light leading to the correlations between the amplitude and phase quadratures of light. Such correlations can be utilized to reduce the fluctuations of the optical field below the SQL~\cite{Fabre,Mancini}. So far, various schemes have been put forward for generating squeezed states either in the light~\cite{lightsqueez} or the mechanical system~\cite{mechsqueez} and, quite recently, squeezing of the optical field has been experimentally realized in optomechanical systems~\cite{Kurn,Painter2013,RegalPRX}. Experimental realization of the squeezed states is also extended to create and stabilize entangled states~\cite{Korppi2018}, which are important for quantum information processing and precision measurements.

In recent years, it has been reported that hybrid atom-assisted optomechanics shows advantages in many aspects~\cite{hybridreview}. To name but a few, atoms enhance the radiation pressure and lead to squeezing of the mechanical mode~\cite{nori}; atoms boost the cooling of the mechanical motion~\cite{atomcooling} and can be utilized to prepare non-Gaussian~\cite{cirac2013} and nonclassical mechanical states by atomic detections~\cite{jie1402}; and the strong coupling between an atomic ensemble and a mechanical oscillator allows to realize quantum control of the oscillator via manipulating the atomic states~\cite{atomOM}. It has also been shown that tripartite entangled stationary states of an atom-cavity-mirror system can be produced~\cite{vitali3,Chiara}. Moreover, nonlocal properties have been investigated in such a tripartite system by the violation of the Mermin-Klyshko inequality~\cite{jie1402}.

In view of the above merits, in this paper we focus on studying optical squeezing in such a hybrid atom-optomechanical system. We explore the atomic effects on the squeezing of the cavity field by considering, e.g., post-selected measurements on the atoms. We show that when the system is at high temperatures, e.g. $T{\sim} 10^2$ K (refer to Sec.~\ref{highT} for the parameters employed), with quadrature fluctuations of light much above the SQL, atomic excitation counting can significantly reduce the noise in one quadrature of optical field without enlarging the noise in the conjugated one. Further reduction can be achieved by increasing the atom-light coupling implemented simply by increasing the number of atoms. However, when the noise is already reduced to a low level close to the SQL, projections no longer work effectively by further increasing the atom-light coupling. The noise approaches the SQL but fails to surpass it. Nevertheless, when the temperature decreases to $T{\sim} 10$ mK which is typically achievable in current experiments, we find the noise of one quadrature is below the SQL without performing any measurements. Significant squeezing of light is observed in the steady state with experimentally feasible parameters. By increasing the atom-light coupling, the squeezing is enhanced with the optimal laser power satisfying the stability condition. Further squeezing below the SQL is obtained by decreasing various losses in the system. We also show that the presence of atoms and excitation counting on the atoms can improve the environment temperature for the appearance of squeezing. The squeezing below the SQL persists for the temperatures of the order K in the atom-optomechanical system. Lastly, we study the nonclassicality of the optical field by making excitation counting on the atomic ensemble, and we observe negative Wigner distributions at high temperatures and the nonclassicality increases with the excitation counting number. Larger effective coupling strength of the system leads to a more squeezed and non-classical state. The output squeezing spectrum and nonclassicality of light is also confirmed.

\section{The system}
\label{system}
\begin{figure}[h]
\includegraphics[width=0.65\linewidth]{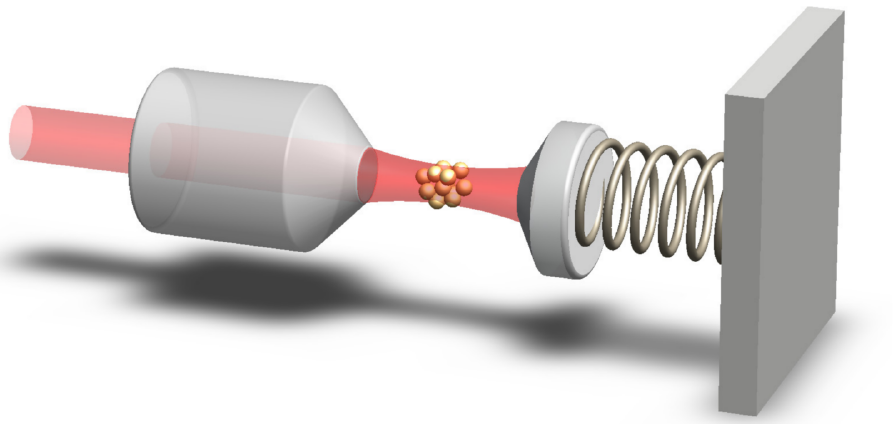}
\caption{Schematic diagram of the system: a cloud of two-level atoms is placed inside a Fabry-Perot cavity with a light movable mirror, which is modelled as a quantum harmonic oscillator. The cavity field mediates the interaction between the atoms and the mechanical oscillator.}
\label{model}
\end{figure}

As shown in Fig.~\ref{model}, we consider a system of an ensemble with $N$ two-level atoms, each with the same eigenfrequency $\omega_a$, placed in a Fabry-Perot cavity with a vibrating end mirror, which is treated as a quantum-mechanical harmonic oscillator with effective mass $m$ and frequency $\omega_m$. The cavity is driven by a pump laser at frequency $\omega_l$. In the unitary picture, without considering any dissipation and decoherence, the Hamiltonian of the system is
\begin{equation}
\begin{split}
H&{=}\hbar \omega_c c^\dagger c+\frac{\hbar}{2}\omega_a S^z+\frac{\hbar}{2}\omega_m (q^2+p^2)\\
&-\hbar\chi c^\dagger c q+\hbar g(S^+c+S^-c^\dagger)+i\hbar\varepsilon(c^\dagger e^{-i\omega_lt}-ce^{i\omega_lt}),
\end{split}
\label{hamil}
\end{equation}
where $\omega_c$ ($\omega_m$) is the frequency of cavity (mechanical) mode, and $c$ ($c^\dagger$) corresponds to the optical annihilation (creation) operator with $[c,c^\dagger ]=1$. The dimensionless mechanical position and momentum operators $q$ and $p$ satisfy $[q,p]=i$. $N$ two-level atoms with the frequency $\omega_a$ compose the atomic ensemble, and each atom can be depicted by the spin-$1/2$ algebra of Pauli matrices $\sigma^{\pm}$ and $\sigma^{z}$. The collective spin operators of the atomic ensemble are defined as $S^{\pm,z}{=}\sum_i\sigma^{\pm,z}_i$ ($i{=}1,2,...,N$), and satisfy the commutation relations $[S^+,S^-]{=}S^z$ and $[S^z,S^{\pm}]=\pm2S^{\pm}$. $\chi$ is the single-photon optomechanical coupling rate given by $\chi=(\omega_c/L)\sqrt{\hbar/m\omega_m}$, where $L$ is the cavity length. The atom-cavity coupling rate $g$ is given by $g=d\sqrt{\omega_c/2\hbar\epsilon_0 V}$, where $V$ is the cavity mode volume, $\epsilon_0$ is the vacuum permittivity and $d$ is the dipole moment of the atomic transition. The driving laser amplitude $\varepsilon$ is related to the pump power $P$ and the cavity decay rate $\kappa$ by $\varepsilon=\sqrt{2P\kappa/\hbar\omega_l}$.

The dynamics of this tripartite system is in principle complicated. Nevertheless, an analytical solution is obtainable for the single atom situation~\cite{favero}. Significant simplification of calculation can be achieved by assuming the low atomic excitation limit, i.e., atoms are all initially prepared in the ground state and the excitation probability of a single atom is small. For highly oriented many atom systems we can use the Holstein-Primakoff approximation~\cite{Holstein}. In this limit, the bosonic annihilation operator $a=S^-/\sqrt{|\langle S^z\rangle|}$ and its Hermitian conjugate $a^\dagger$ can describe the dynamics of the atomic polarization, which satisfy the bosonic commutation relation $[a,a^\dagger]{=}1$. In the reference frame rotating at the laser frequency $\omega_l$, the quantum Langevin equations accounting for the dynamics of the system can be written as
\begin{equation}
\begin{split}
\dot{q}&=\omega_m p,\\
\dot{p}&=-\omega_m q-\gamma_m p+\chi c^\dagger c+\xi,\\
\dot{c}&=-(\kappa+i\Delta_c)c+i\chi cq-ig_Na+\varepsilon+\sqrt{2\kappa}c_{in},\\
\dot{a}&=-(\gamma_a+i\Delta_a)a-ig_Nc+\sqrt{2\gamma_a}a_{in},\\
\end{split}
\end{equation}
where $\Delta_c{=}\omega_c{-}\omega_l$ ($\Delta_a{=}\omega_a{-}\omega_l$) is the cavity-pump (atom-pump) detuning with respect to the driving light, and $\gamma_m$ ($\gamma_a$) is the  mechanical (atomic) decay rate. The coupling between the cavity and the collective atomic modes is denoted by $g_N{=}\sqrt{N}g$. The operators $\{\xi,c_{in},a_{in}\}$ account for the zero-mean input noises affecting the mirror, optical and atomic modes, respectively. The Langevin force operator $\xi$, which models the effects of the mechanical Brownian motion, has a non-Markovian correlation~\cite{DVitali}
\begin{equation}
\langle\xi(t)\xi(t')\rangle{=}\frac{\gamma_m}{2\pi\omega_m}\int\omega \text{e}^{-i\omega(t-t')}[\text{coth}(\frac{\hbar\omega}{2k_BT})+1]\text{d}\omega,
\end{equation}
where $k_B$ is the Boltzmann constant and $T$ is the temperature of the phononic environment. In the limit of a high mechanical quality factor, the above correlation becomes delta-correlated ~\cite{deltafunction}. Since the cavity and atomic modes are both prepared in coherent states, the only nonvanishing correlation functions of $c_{in}$ and $a_{in}$ are $\langle c_{in}(t)c_{in}^{\dag}(t')\rangle{=}\langle a_{in}(t)a^{\dag}_{in}(t')\rangle{=}\delta(t-t')$~\cite{noisebook}.

The degree of noise reduction in the system depends heavily on the optomechanical coupling strength. To achieve this, we assume the cavity is strongly pumped, i.e. $|\alpha_s|\gg 1$, where $\alpha_s$ is the amplitude of the steady-state cavity field, which can be acquired by solving the nonlinear equation $\alpha_s[\kappa+i\Delta_c-i\chi^2|\alpha_s|^2/\omega_m+g^2_N/(\gamma_a+i\Delta_a)]{=}\varepsilon$. Since we are interested in noise reduction in the quadratures of optical field at the steady state, we thus analyze the fluctuations of the system operators around their steady state values and linearize the dynamics by writing the quadrature operators $O{=}(q,p,X,Y,x,y)^{\rm T}$ as $O_i{\simeq}\langle O_i\rangle{+}\delta O_i$, with $\langle O_i\rangle$ the mean value of each operator and $\delta O_i$ the corresponding fluctuation, where we introduced $X{=}(c^\dag{+}c)/\sqrt2$, $Y{=}i(c^\dag{-}c)/\sqrt2$, and $x{=}(a^\dag{+}a)/\sqrt2$, $y{=}i(a^\dag{-}a)/\sqrt2$ the position- and momentum-like operators of the optical and atomic modes, respectively. In such a way, the dynamics of the system takes a linear form that simplifies the cumbersome calculation. The resulting evolution equation for the fluctuation operators $\delta O{=}(\delta q,\delta p,\delta X,\delta Y,\delta x,\delta y)^{\rm T}$ is
\begin{equation}
\delta{\dot O}={\bf K}\delta O+{\bf n},
\label{QLE}
\end{equation}
where the drift matrix {\bf K} is given by
\begin{equation}
{\bf K}=
\begin{pmatrix}
0 & \omega_m & 0 & 0 & 0 & 0 \\
-\omega_m & -\gamma_m & \chi_{e\!f\!f}  & 0 & 0 & 0 \\
0 & 0 & -\kappa & \tilde{\Delta}_c & 0 & g_N \\
\chi_{e\!f\!f} & 0 & -\tilde{\Delta}_c & -\kappa & -g_N & 0 \\
0 & 0 & 0 & g_N & -\gamma_a & \Delta_a \\
0 & 0 & -g_N & 0 & -\Delta_a & -\gamma_a \\
\end{pmatrix},
\label{drift}
\end{equation}
with the effective optomechanical coupling $\chi_{e\!f\!f}{=}\sqrt{2}\chi\alpha_s$ (by choosing a phase reference, $\alpha_s$ can be taken as real), and the effective cavity detuning $\tilde{\Delta}_c=\Delta_c{-}\chi^2_{e\!f\!f}/2\omega_m$. The vector of noise operators {\bf n} writes in the form of ${\bf n}=(0,\xi, \sqrt{2\kappa}X_{in}, \sqrt{2\kappa}Y_{in}, \sqrt{2\gamma_a}x_{in}, \sqrt{2\gamma_a}y_{in} )^{\rm T}$, where we introduced $X_{in}{=}(c^\dag_{in}{+}c_{in})/\sqrt2$, $Y_{in}{=}i(c^\dag_{in}{-}c_{in})/\sqrt2$, and $x_{in}{=}(a^\dag_{in}{+}a_{in})/\sqrt2$, $y_{in}{=}i(a^\dag_{in}{-}a_{in})/\sqrt2$. Equation~(\ref{QLE}) can be solved directly in the frequency domain by taking the Fourier transform of the equation. The correlation function of any pair of fluctuation operators is then obtained by
\begin{equation}
V_{ij}=\frac{1}{4\pi^2}\iint\!\text{d}\omega\text{d}\Omega e^{-i(\omega+\Omega)t}V_{ij}(\omega,\Omega),
\label{correlations}
\end{equation}
with $V_{ij}(\omega,\Omega){=}\langle\{v_j(\omega),v_k(\Omega)\}\rangle/2 ~(i,j{=}1,..,6)$ the frequency-domain correlation function between elements $i$ and $j$ of $v(\omega)=(\delta q(\omega), \delta p(\omega),\delta X(\omega),\delta Y(\omega),\delta x(\omega),\delta y(\omega) )$. Our hybrid optomechanical system is fully determined by the $6\times6$ covariance matrix (CM) ${\bm \sigma}$ with elements defined in Eq.~\eqref{correlations}. Being a physical state, ${\bm \sigma}$ should satisfy the Heisenberg-Robertson uncertainty principle ${\bm\sigma}+i\Omega_3/2\ge0$~\cite{SimonMukunda} with $\Omega_3{=}\oplus^3_{j=1}i\sigma_y$ the so-called symplectic matrix and $\sigma_y$ the $y$-Pauli matrix. The system is stable when all the eigenvalues of the drift matrix {\bf K} have negative real parts~\cite{stable}. In what follows, all the results to be discussed are guaranteed to meet this stability condition.

\section{noise reduction by excitation counting on atoms}
\label{highT}

In this Section, we focus on the quantum fluctuations of optical quadratures and provide a strategy to strongly suppress the noises based on the post-selected measurements, e.g. excitation counting, on the atoms. This is particularly powerful when the system is at high temperatures with large noises because it reduces one quadrature's fluctuation close to the SQL without increasing the noise in the other one. The fluctuations $\delta X$ and $\delta Y$ are subject to the Heisenberg uncertainty relation. For vacuum or coherent states, we have $\langle\delta X^2\rangle=\langle\delta Y^2\rangle=1/2$. The fluctuations below 1/2 indicates that the SQL is surpassed and, correspondingly, the state is called a squeezed state.

We first discuss the situations when the system is at high temperatures. In this case, the quadratures of the system have large fluctuations, i.e., $\langle\delta X^2\rangle$ or $\langle\delta Y^2\rangle\gg 1/2$. In the following, we show that with excitation counting on the atoms, $\langle\delta X^2\rangle$ can be reduced to approach the SQL, i.e., $\langle\delta X^2\rangle\sim1/2$.
\begin{figure}[t]
{\bf (a)}\hskip3.7cm{\bf (b)}
\includegraphics[width=0.45\linewidth]{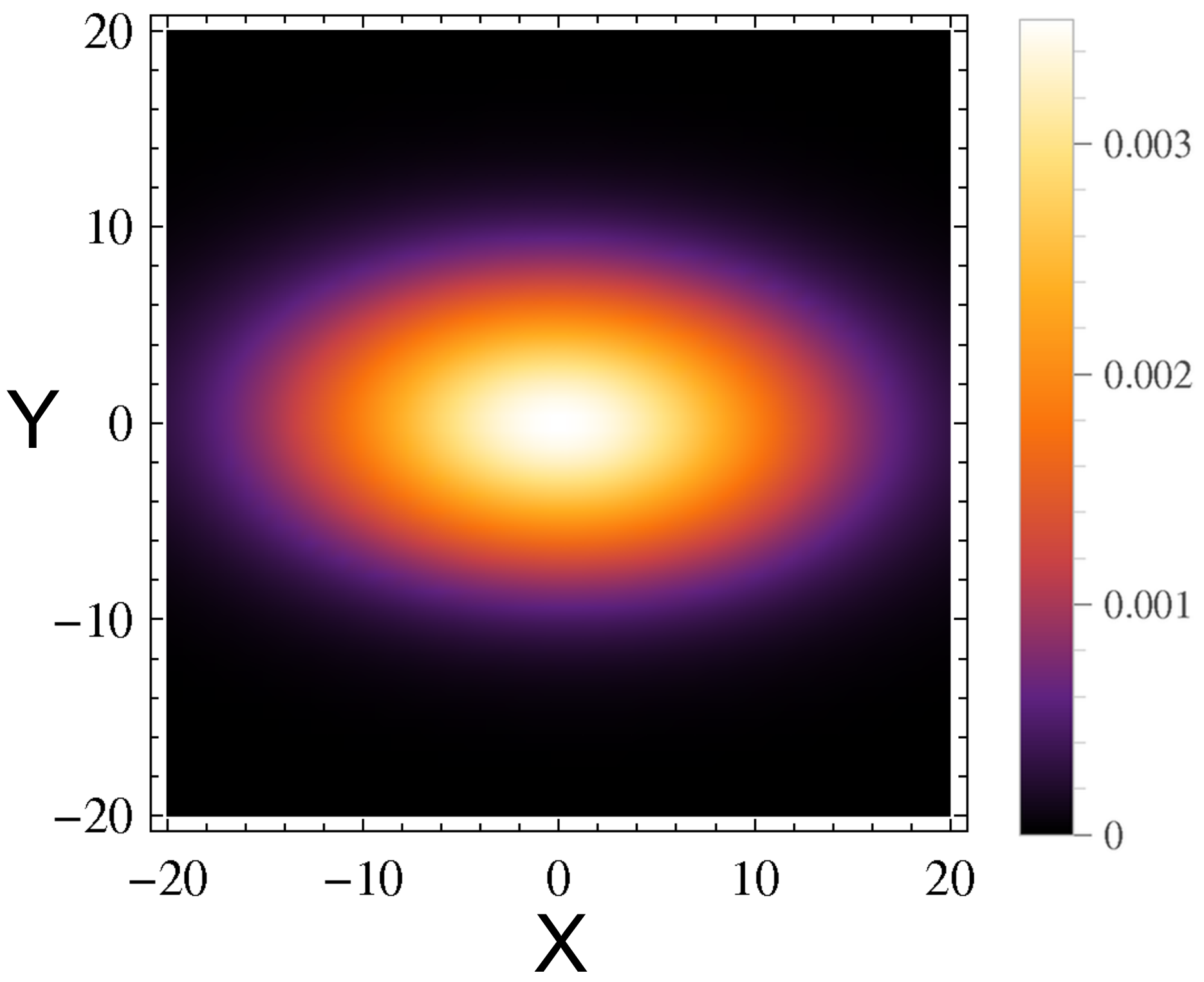}~~~\includegraphics[width=0.45\linewidth]{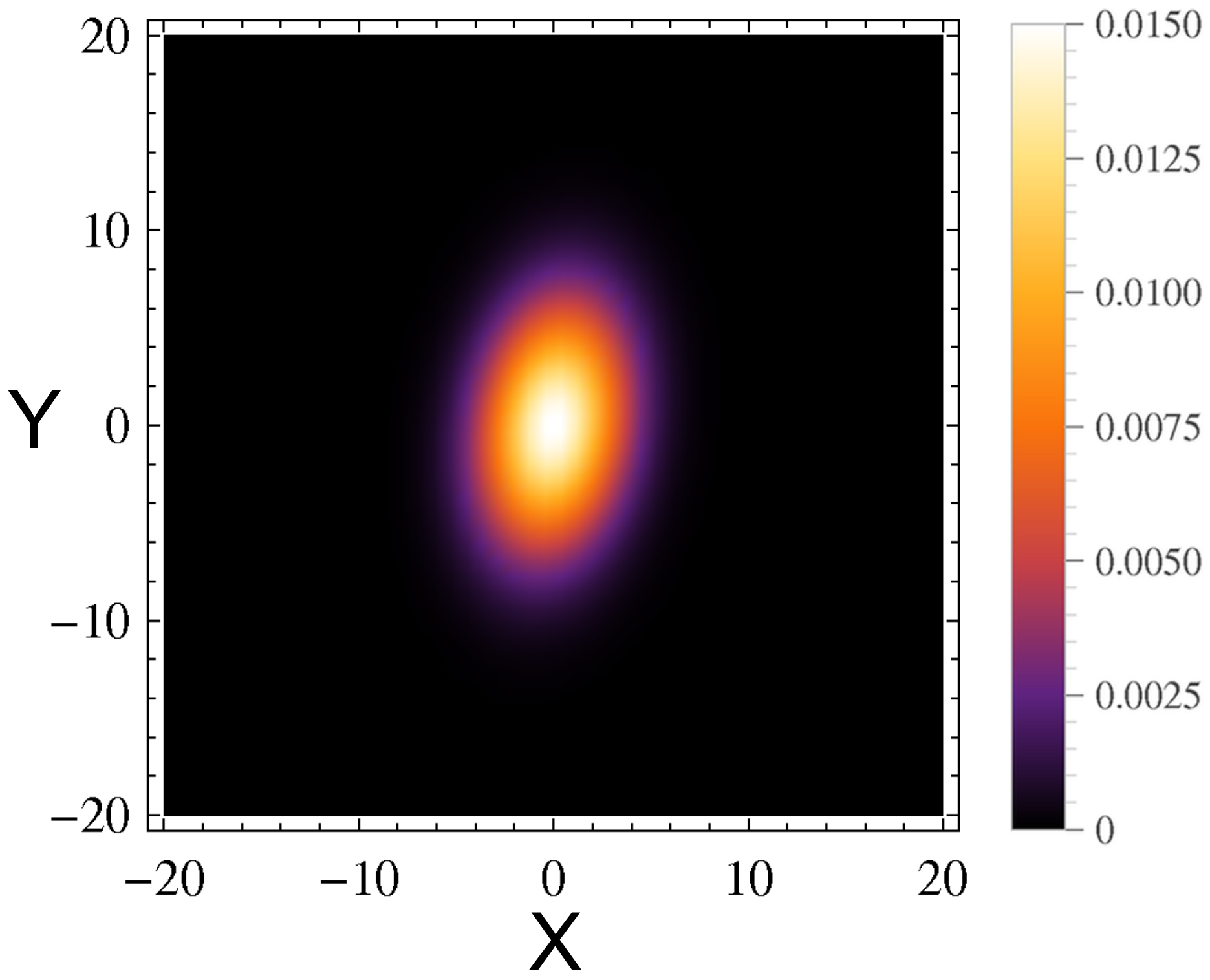}\\
{\bf (c)}\hskip3.7cm{\bf (d)}
~~~\includegraphics[width=0.45\linewidth]{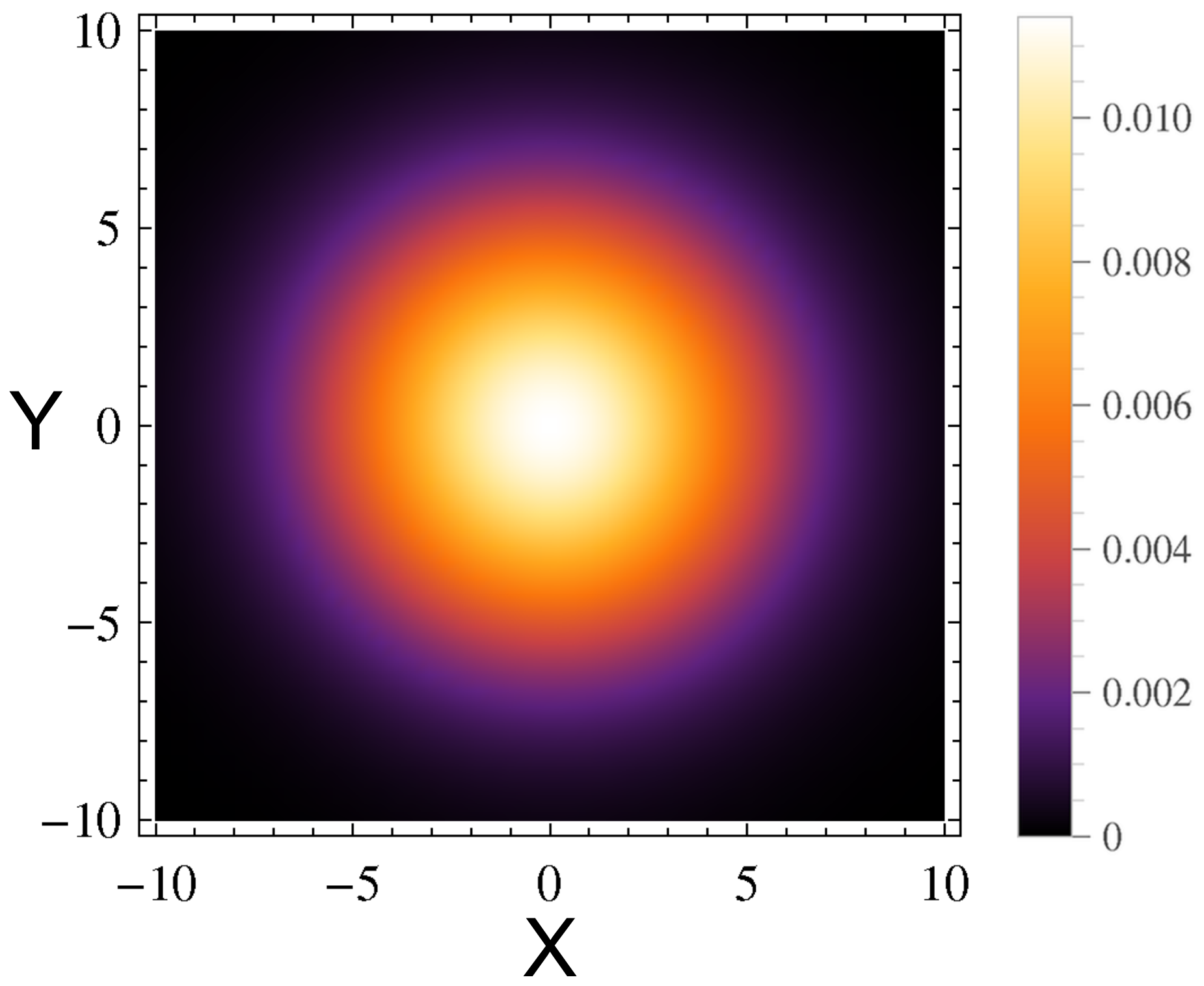}~~~\includegraphics[width=0.45\linewidth]{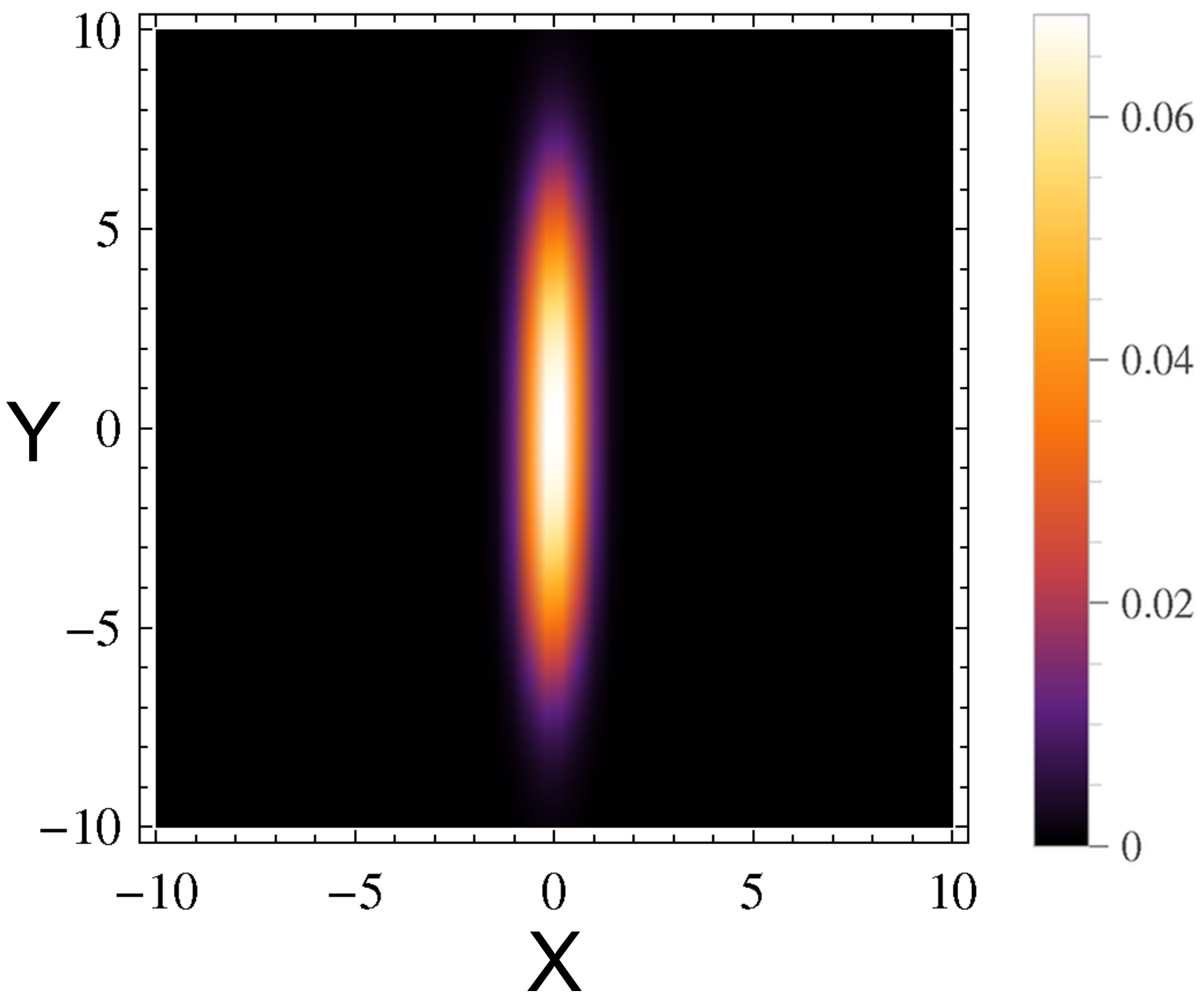}
\caption{Wigner function of the cavity field without [{\bf (a)} and {\bf (c)}] and/or with [{\bf (b)} and {\bf (d)}] excitation counting $\Pi{=}|1\rangle_a\langle 1|$ on the atomic state. In {\bf (a)} and {\bf (b)} the atom-light coupling takes $g_N=2\times10^8$ Hz, and {\bf (c)} and {\bf (d)} take a larger value, $g_N=4\times10^8$ Hz. Other parameters see text for details. {\bf (a)} $\langle\delta X^2\rangle=164.65$, $\langle\delta Y^2\rangle=48.72$; {\bf (b)} $\langle\delta X^2\rangle=12.52$, $\langle\delta Y^2\rangle=34.45$; {\bf (c)} $\langle\delta X^2\rangle=27.93$, $\langle\delta Y^2\rangle=27.64$; {\bf (d)} $\langle\delta X^2\rangle=0.54$, $\langle\delta Y^2\rangle=16.86$. In {\bf (d)}  $\langle\delta X^2\rangle$ is reduced to just above the SQL.}
\label{wignerplot}
\end{figure}
Having acquired the CM of the system, its characteristic function can be directly achieved by~\cite{parisbook}
\begin{equation}
\zeta(O)={\rm exp}(-O^{\rm T}{\bm \sigma}\,O).
\label{66cm}
\end{equation}
For the cavity field subsystem, similarly, we have $\zeta(X,Y)={\rm exp}[-(X,Y)\,{\bm \sigma_c}\,(X,Y)^{\rm T}]$, where ${\bm \sigma_c}$ is the CM associated with the light only. After the replacements $q={\rm Re}[\alpha],\, p={\rm Im}[\alpha];\, X={\rm Re}[\beta],\, Y={\rm Im}[\beta]$;\, and $x={\rm Re}[\gamma],\, y={\rm Im}[\gamma]$ with amplitude $\{\alpha,\beta,\gamma\}\in\mathbb{C}$, the characteristic function is rewritten as $\zeta(\alpha,\beta,\gamma)$. This gives us access to the density matrix of the system~\cite{glauber}:
\begin{equation}
\rho_{mca}=\frac{1}{\pi^3}\iiint \text{d}^2\alpha\,\text{d}^2\beta\,\text{d}^2\gamma\,\zeta(\alpha,\beta,\gamma)D_m(-\alpha)D_c(-\beta)D_a(-\gamma),
\label{rhosystem}
\end{equation}
where $D_j(\mu)$ is the displacement operator of mode $j=m,c,a$~\cite{Wallsbook}. Now we implement excitation counting on the atomic state. For low number excitation counting, this can be performed by utilizing a time-resolved and position-sensitive detector (microchannel plate and delay-line anode)~\cite{Jeltes}. For large number excitation counting, Rydberg excitation and its detection are helpful for the measurement. In this case, principal quantum numbers is associated with the number of ground-state atoms in each Rydberg excitation, and the Rydberg excitations are field ionized using an approximately linear field ionization ramp. Large number atom counting can be carried out by using microchannel plate detection and Rydberg states~\cite{atomcount}. This gives rise to the following density matrix for the conditional state of the cavity field:
\begin{equation}
\rho_c={\rm Tr}_{m,a}\left[\Pi \, \rho_{mca}\right]/{\rm Tr}_{m,c,a}\left[\Pi \, \rho_{mca}\right],
\label{rhomirror}
\end{equation}
where operator $\Pi{=}|s\rangle_a\langle s|$ denotes excitation counting on the atoms and the denominator is a normalization constant. Note that due to the low excitation assumption, $s$ should be a small integer. As we will see, actually $s=1$, i.e., projecting the atoms onto a single excitation state, is sufficient to strongly suppress the noise of the cavity field. Once having the density matrix of the cavity field, one then obtains its characteristic function by
\begin{equation}
\zeta(\lambda)={\rm Tr}[D_c(\lambda)\,\rho_c],
\end{equation}
with $\lambda\in\mathbb{C}$. The analytical expression of $\zeta(\lambda)$ is provided in the Appendix. Since $\zeta(\lambda)$ contains the information of fluctuations $\langle\delta X^2\rangle$ and $\langle\delta Y^2\rangle$, the degree of the noise reduction can then be calculated by comparing with the initial $\zeta(\beta)$ before the measurement implemented. From $\zeta(\lambda)$, one can also derive the Wigner function of the cavity field by $W(\lambda)={\cal F}[\zeta(\lambda)]$, where ${\cal F}[\cdot]$ denotes taking the Fourier transform.

Figure~\ref{wignerplot} shows the Wigner function of the cavity field with and/or without excitation counting on the atoms for two values of $g_N$, in which we employed experimentally feasible parameters~\cite{Painter2013}: $m{=}2\times10^{-13}$ g, $\omega_m/2\pi{=}10^7$ Hz, a lower Q factor $Q{=}\omega_m/\gamma_m{=}10^4$, and a higher temperature $T{=}100$ K; laser power $P{=}2.36$ mW at $\lambda_l{=}1540$ nm, cavity length $L{=}1$ mm and decay rate $\kappa/2\pi=2\times10^6$ Hz; atomic decay rate $\gamma_a/2\pi{=}2\times10^5$ Hz and $g_N=2\times10^8$ Hz in Fig.~\ref{wignerplot} {\bf (a)} and {\bf (b)}, and $g_N=4\times10^8$ Hz in {\bf (c)} and {\bf (d)}. In view of a cloud of atoms ($N>10^7$) trapped in the cavity~\cite{McConnell}, we have considered a larger cavity and the power and the decay rate are accordingly adjusted to guarantee the stability of the system. We take optimal detunings as $\tilde{\Delta}_c\!\sim\!0$ and $\Delta_a\!\sim\!-\omega_m$, which will be expounded in the next Section. As shown in Fig.~\ref{wignerplot}, with a sizeable atom-light coupling $g_N$, atomic counting can strongly suppress $\langle\delta X^2\rangle$ to just above the SQL. However, by further increasing $g_N$, measurements fail to reduce the niose below the SQL but push it very close to the limit. For a very large $g_N=10^9$ Hz, $\langle\delta X^2\rangle$ is reduced to 0.503 still above the SQL. Moreover, non-classical state exhibiting negative Wigner function can be generated by increasing the number of excitation counting, even at the high temperature $T\sim100$ K. The negativity of the Winger function robust against the effects of the environment affecting the atom-optomechanical system, and should be taken as a sort of witness for nonclassicality. In section IV, we will stress the nonclassicality in detail.

\section{quantum noise below the SQL}
\label{belowSQL}

We now show that, by lowering the temperature, quadrature squeezing of light below the SQL can be generated in this hybrid atom-optomechanical system and measurements on atoms play an active role in enhancing the squeezing.

\begin{figure}[t]
\hskip0.25cm\includegraphics[width=0.85\linewidth]{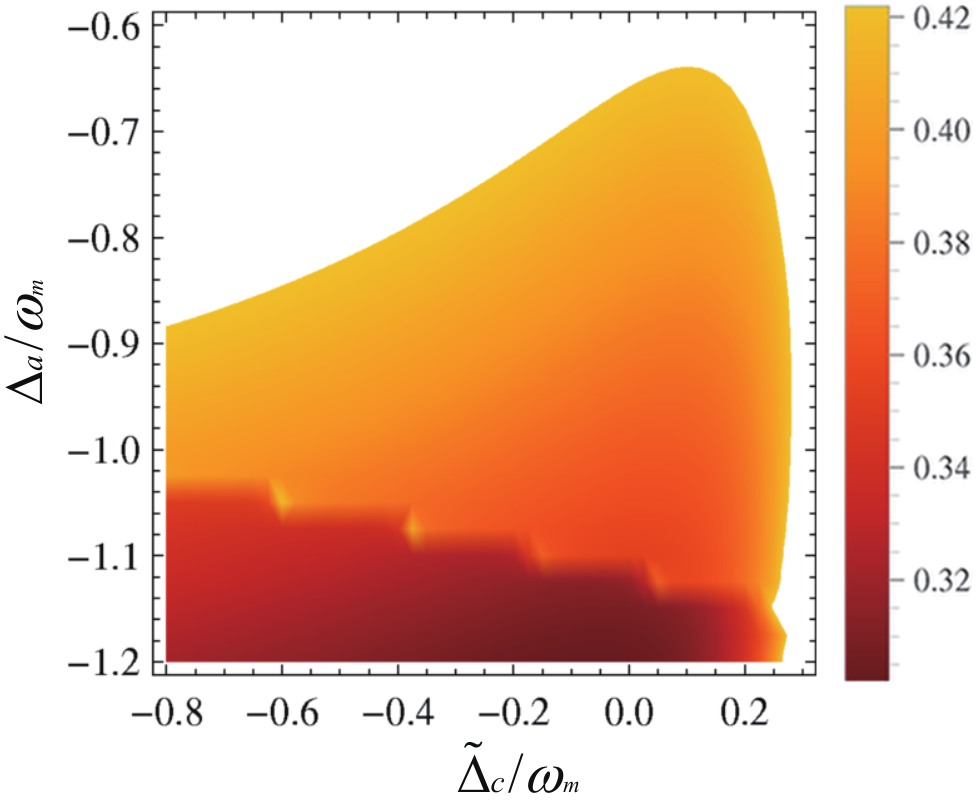}
\caption{ $\langle\delta Y^2\rangle$ as a funtion of detunings $\tilde{\Delta}_c/\omega_m$ and $\Delta_a/\omega_m$. The darker area at the bottom represents the states that are not stable and thus should not be considered for our aim. The blank area denotes states with $\langle\delta Y^2\rangle{>}0.422$. The same conditions as in Fig.~\ref{wignerplot} {\bf (a)} but for $T{=}0.01$ K.}
\label{detunings}
\end{figure}

Determining the optimal values of the parameters, especially for the detunings, is the key to generate squeezing in our system. In Fig.~\ref{detunings}, we show the fluctuation $\langle\delta Y^2\rangle$ of light as a function of the two detunings $\tilde{\Delta}_c$ and $\Delta_a$ with the same parameters as in Fig.~\ref{wignerplot} {\bf (a)} but for a lower temperature $T{=}10$ mK and without measurements on atoms. It shows that the optimal detunings for the squeezing of $\langle\delta Y^2\rangle$ at the steady state are: $\tilde{\Delta}_c\sim0$ and $\Delta_a\sim-\omega_m$ (refer to Refs.~\cite{vitali3,jie1402} for different optimal detunings for the entanglement and nonlocality in such a tripartite system). Largest squeezing $\langle\delta Y^2\rangle\sim0.355$ (29$\%$ below the SQL) is found close to the unstable regime~\cite{taylor}.

\begin{figure}[t]
\includegraphics[width=0.9\linewidth]{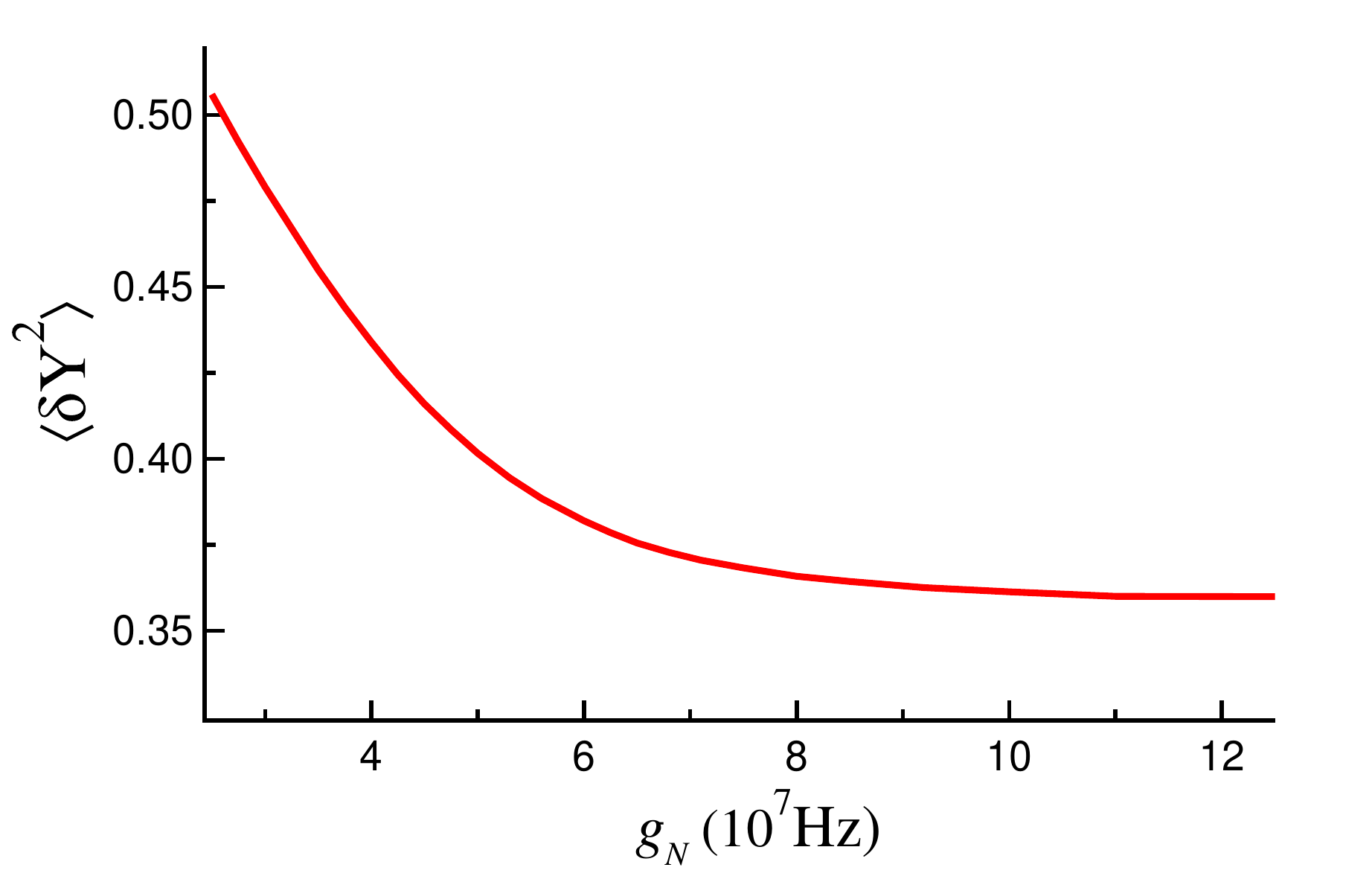}
\caption{$\langle\delta Y^2\rangle$ as a function of $g_N$. The same conditions as in Fig.~\ref{detunings} but with $\tilde{\Delta}_c{=}0$ and $\Delta_a{=}-\omega_m$. For a certain value of $g_N$, $\langle\delta Y^2\rangle$ is optimized by adjusting the power $P$ satisfying the stability.}
\label{Y2gN}
\end{figure}

For a given $\tilde{\Delta}_c$, the nonlinear equation for the amplitude of the steady-state cavity field $\alpha_s$ can be solved by adjusting the cavity-laser detuning $\Delta_c$, and is given by
\begin{equation}
\alpha_s{=}\varepsilon/[\kappa+i\tilde{\Delta}_c+g^2_N/(\gamma_a+i\Delta_a)],
\end{equation}
from which we see that the laser power $P$ (included in $\varepsilon$) and the atom-light coupling $g_N$ mainly determine the value of $\alpha_s$ and thus determine the optomechanical coupling $\chi_{e\!f\!f}$, which plays a key role in light squeezing in our system. Numerical calculations show that for a given $g_N$, $P$ close to the maximum value allowed by the stability condition yields optimal squeezing of $\langle\delta Y^2\rangle$. The degree of squeezing is enhanced till to a saturated value as $g_N$ increases, as shown in Fig.~\ref{Y2gN}. Since the {\it optimal} power $P$ and $\chi_{e\!f\!f}$ increase as $g_N$ grows, one would suspect this enhancement is solely caused by the increase of $\chi_{e\!f\!f}$. This is excluded by the fact that when $g_N$ changes from $10^8$ Hz to $2\times10^8$ Hz (not shown), the optimal $\chi_{e\!f\!f}$ varies from $9.07\times10^7$ Hz to $1.8\times10^8$ Hz, while $\langle\delta Y^2\rangle$ is almost unchanged. Further squeezing would occur by decreasing various losses in the system. For example in Fig.~\ref{Y2gN}, if we take $g_N{=}1.25\times10^8$ Hz and decrease $\gamma_m$ and $\gamma_a$ to one percent of their original values, i.e.,
$\gamma_m/2\pi{=}10$ Hz and $\gamma_a/2\pi{=}2\times10^3$ Hz, $\langle\delta Y^2\rangle$ is reduced to 0.325, i.e., $35\%$ below the SQL. The excitation counting $\Pi{=}|1\rangle_a\langle 1|$ on the atomic state slightly improves the squeezing of $\langle\delta Y^2\rangle$ to $36\%$ below the SQL.

The squeezing in the amplitude quadrature $\langle\delta X^2\rangle$ of light is optimized for $\tilde{\Delta}_c\sim\omega_m$ and $\Delta_a\sim-\omega_m$, i.e. the cavity is resonant with the anti-Stokes sideband of the laser, while the atoms are resonant with the Stokes sideband. The optimal detunings for the entanglement and nonlocality in such a tripartite system are the same~\cite{vitali3,jie1402}. The maximum squeezing of $\langle\delta X^2\rangle$ could be achieved with experimentally feasible parameters~\cite{Kurn,hybridreview}: $\omega_m/2\pi{=}4.92\times10^5$ Hz, $\gamma_m/2\pi{=}0.1$ Hz, and lower temperature $T{=}0.01$ K. The laser power $P$ and the coupling $\chi_{e\!f\!f}$ are increased to 90 mW and $7.15\times10^6$ Hz. In comparison with the losses mentioned above, the cavity and atomic decay rate are decreased to $\kappa/2\pi=8\times10^3$ Hz and $\gamma_a/2\pi{=}8\times10^2$ Hz. In this high-finesse cavity ($F>10^5$), we use a longer cavity $L{=}20$ mm and decrease the number of trapped atoms ($N\sim10^6$, corresponding to $g_N{=}1.11\times10^7$ Hz) to make the experiment more feasible. The optimal squeezing of $\langle\delta X^2\rangle\sim0.317$ is found in the case, i.e., $37\%$ below the SQL. The single excitation counting enhances the squeezing to $40\%$ below the SQL, which plays a stronger role compared with the case of phase quadrature $\langle\delta Y^2\rangle$.

\begin{figure}[htb]
\includegraphics[width=0.9\linewidth]{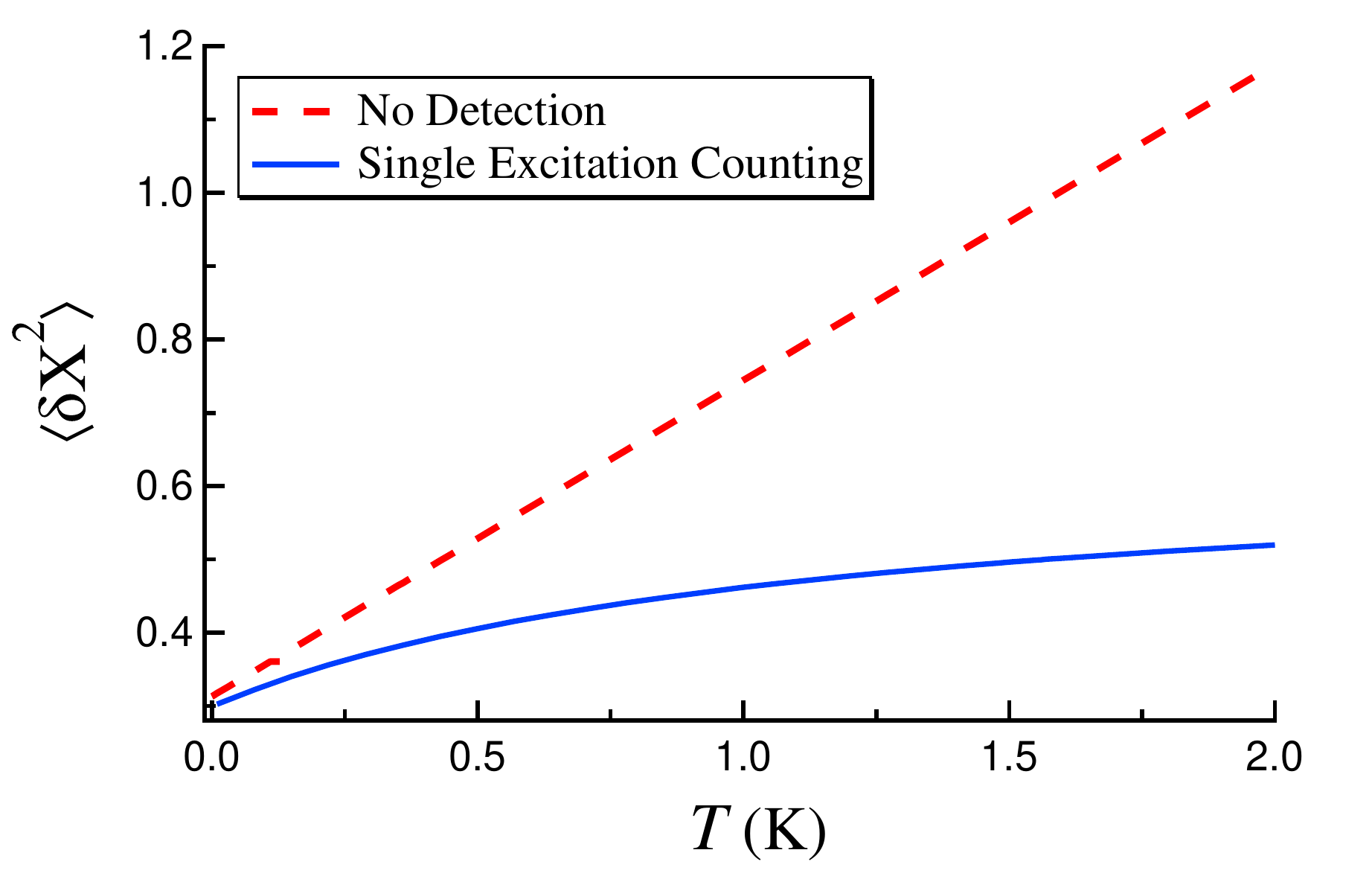}
\caption{$\langle\delta X^2\rangle$ as a function of $T$. The parameters see text for details. The red dashed line refers to the presence of atoms without detection; the blue solid curve refers to single excitation counting $\Pi{=}|1\rangle_a\langle 1|$ on the atomic state.}
\label{X2T}
\end{figure}

The robustness of the squeezing $\langle\delta X^2\rangle$ dependent on the environment temperature $T$ is shown in Fig.~\ref{X2T}. The parameters are the same as those used for obtaining the optimal squeezing of $\langle\delta X^2\rangle$. With the help of the atoms, the squeezing of $\langle\delta X^2\rangle$ persists for environmental temperatures above 0.43 K without any post-selected measurements (see the red dashed line in Fig.~\ref{X2T}), which is two orders of magnitude larger than the temperature for squeezing in the absence of atoms. The environment temperature corresponding to the appearance of squeezing is further increased to 1.56 K with the single excitation
counting $\Pi{=}|1\rangle_a\langle 1|$ on the atoms (blue solid curve shown in Fig.~\ref{X2T}). Hence, the presence of atoms and excitation counting on the atoms can reduce the requirements regarding thermal noise from environment, and enhance the squeezing at relatively high temperatures compared to the case without atoms.

\section{Inducing negativity of the Wigner function}
\label{output}
\begin{figure}[h]
{\bf (a)}
\includegraphics[width=0.85\linewidth]{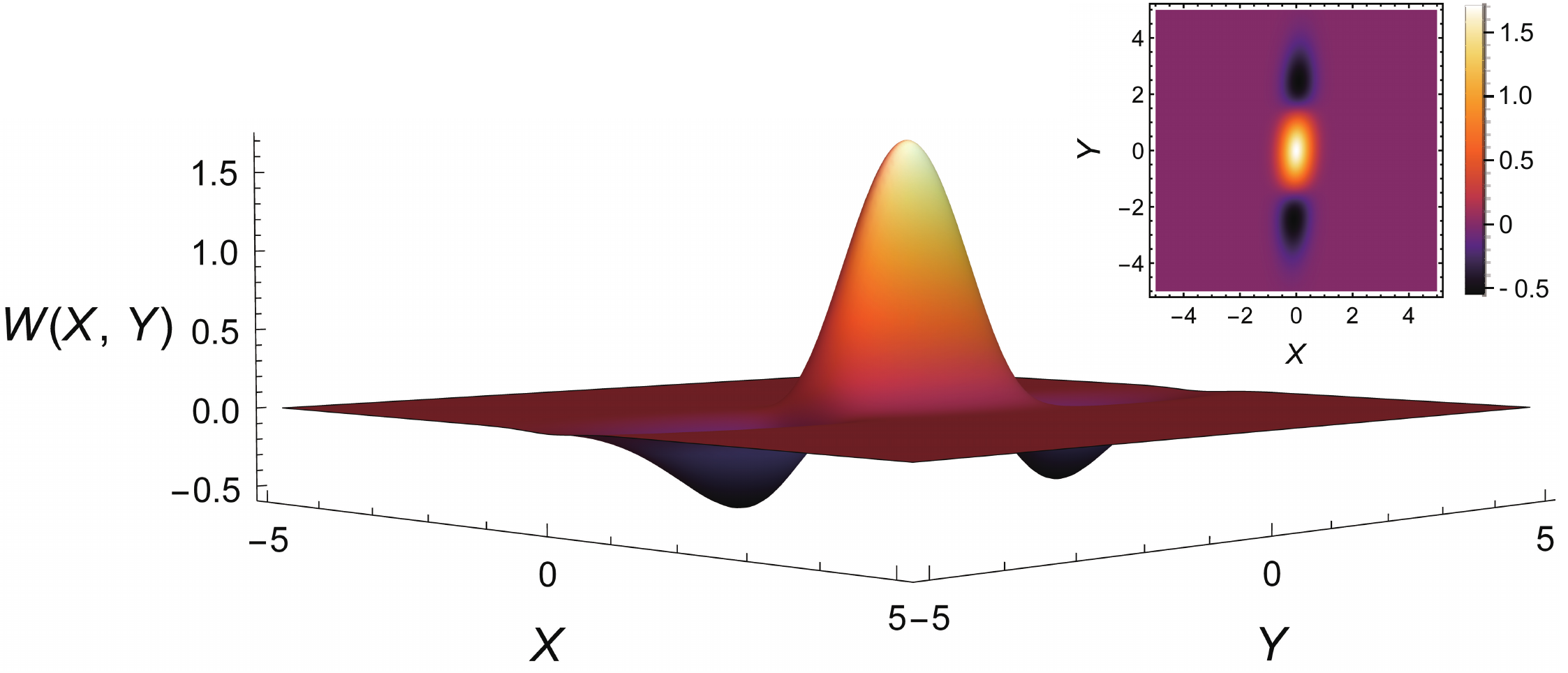}
\vskip0.5cm{\bf (b)}
\includegraphics[width=0.85\linewidth]{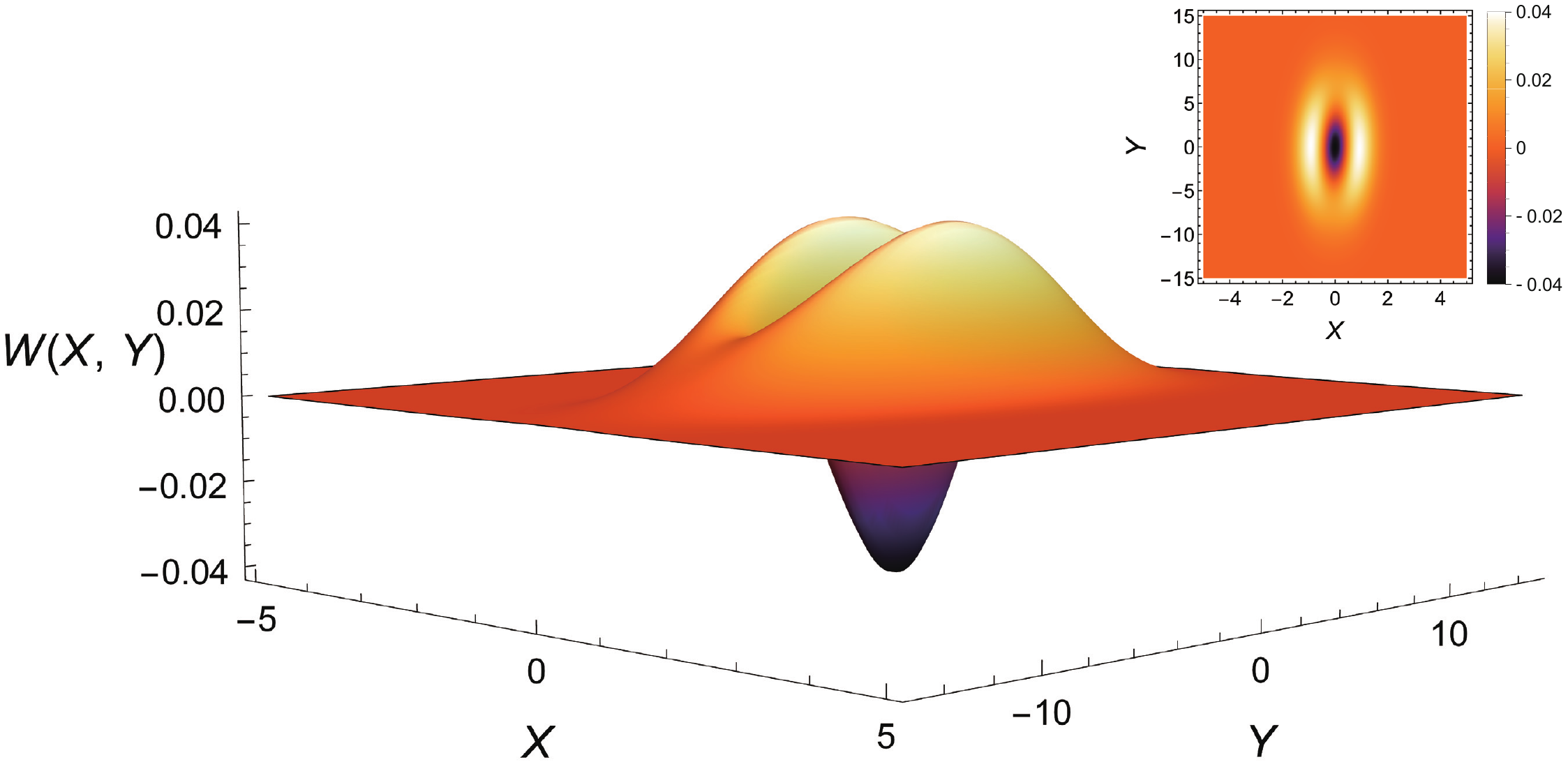}
\caption{Wigner function of the optical field with excitation counting $\Pi{=}|2\rangle_a\langle 2|$ on the atomic state. See the text for details of the parameters. {\bf (a)} $T=10$ mK; {\bf (b)} $T=100$ K.}
\label{intrawigner}
\end{figure}

While the above sections are focused on the squeezing of light, in this section, we concentrate on the effectiveness of the post-selected measurement in inducing negative Wigner functions of the light~\cite{Kenfack}. Due to the Gaussian nature of the linearized optomechanical interaction, negativity of the Wigner function will not be induced through Gaussian measurements~\cite{Li}. Using conditional non-Gaussian measurements, e.g. excitation counting on the atoms, the nonclassicality of light with a negative Wigner function is strongly induced. As shown in Fig.~\ref{intrawigner}, we present the Wigner distribution of the light with excitation counting $\Pi{=}|2\rangle_a\langle 2|$ on the atomic state. The parameters are the same as those used for achieving the largest squeezing of $\langle \delta X^2\rangle$ in the last section. In Fig.~\ref{intrawigner} {\bf (a)}, it shows that a negative Wigner function is effectively induced through excitation counting on the atoms. It should be noted that single-excitation counting $\Pi{=}|1\rangle_a\langle 1|$ on the atoms is inefficient to induce a negative Wigner function, while two-excitation counting can. This is because a larger excitation counting number yields stronger nonclassicality of the light~\cite{Li}. Quite noticeably, even classical pumping is sufficient to induce the negativity of the Wigner function. The negativity is quite robust against temperature: by increasing the temperature up to $T\sim100$ K, the negativity is still maintained, as shown in Fig.~\ref{intrawigner} {\bf (b)}. The robustness is also found against the mechanical and atomic decay rate: the Wigner
function sustains its negativity as $\gamma_m$ increases by four orders of magnitude and as $\gamma_a$ increases by one order of magnitude.

\begin{figure}[htb]
\includegraphics[width=0.95\linewidth]{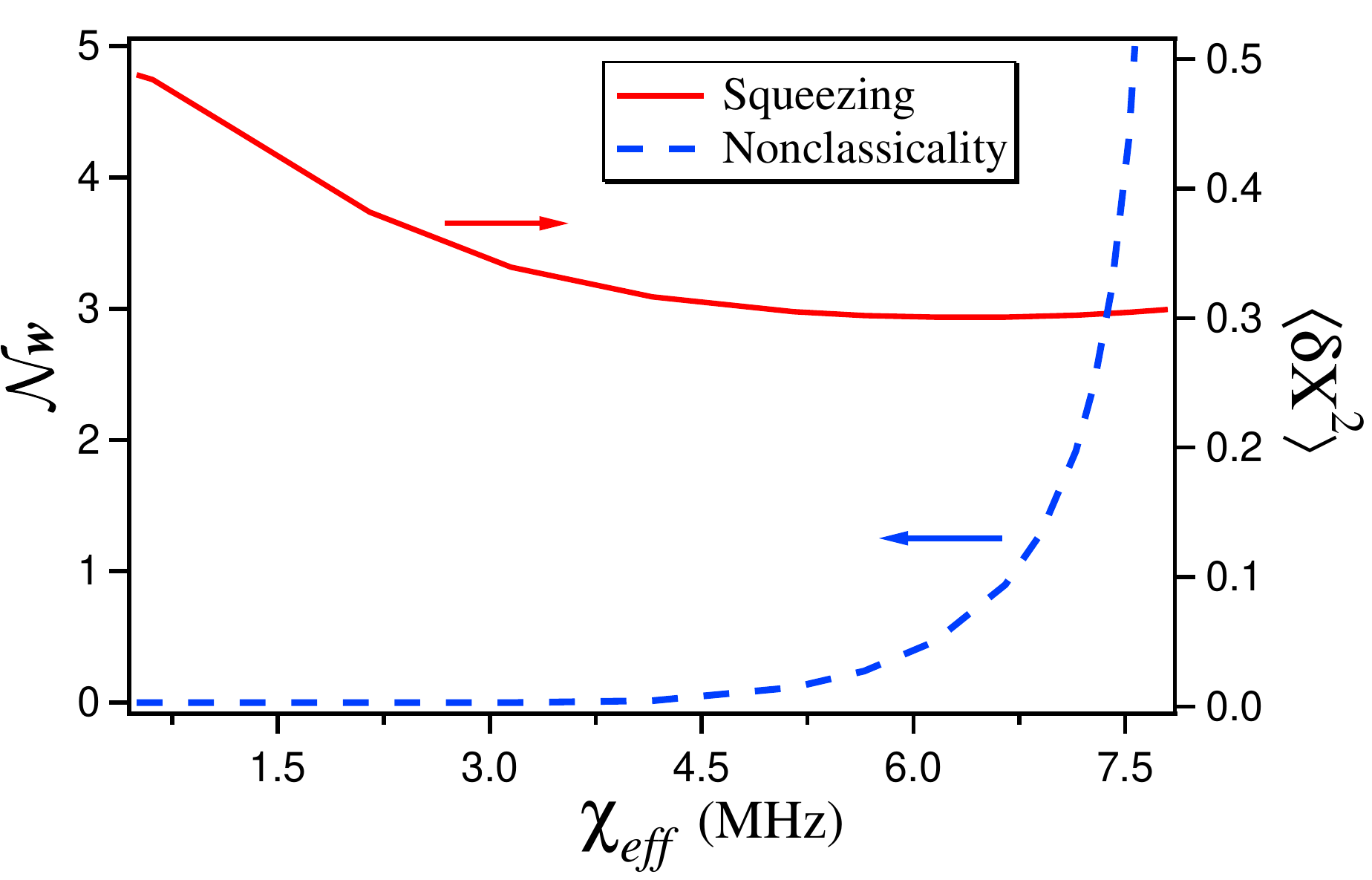}
\caption{Optical nonclassicality ${\mathcal{N}}_{w}$ and squeezing $\langle\delta X^2\rangle$ versus effective optomechanical coupling $\chi_{e\!f\!f}$ with excitation counting $\Pi{=}|2\rangle_a\langle 2|$ on the atomic state. The parameters are the same as those used in Fig.~\ref{intrawigner} {\bf (a)}.}
\label{NwS}
\end{figure}

We also show that the induced nonclassicality and the enhanced squeezing is tightly associated with the effective optomechanical coupling $\chi_{e\!f\!f}$: using excitation counting $\Pi{=}|2\rangle_a\langle 2|$ on the atoms, stronger nonclassicality and suqeezing is generated and maintained as the coupling increases, as shown in Fig.~\ref{NwS}. We adopt the nagative values of the Wigner function ${\mathcal{N}}_{w}$ to quantify the nonclassicality of the light.The ${\mathcal{N}}_{w}$ is defined as ${\mathcal{N}}_{w}=\left\vert \int_{\Phi }W(\delta )d^{2}\delta \right\vert $, where $\Phi $ is the negative region of the Wigner distribution in phase space.

\begin{figure}[htb]
\includegraphics[width=0.95\linewidth]{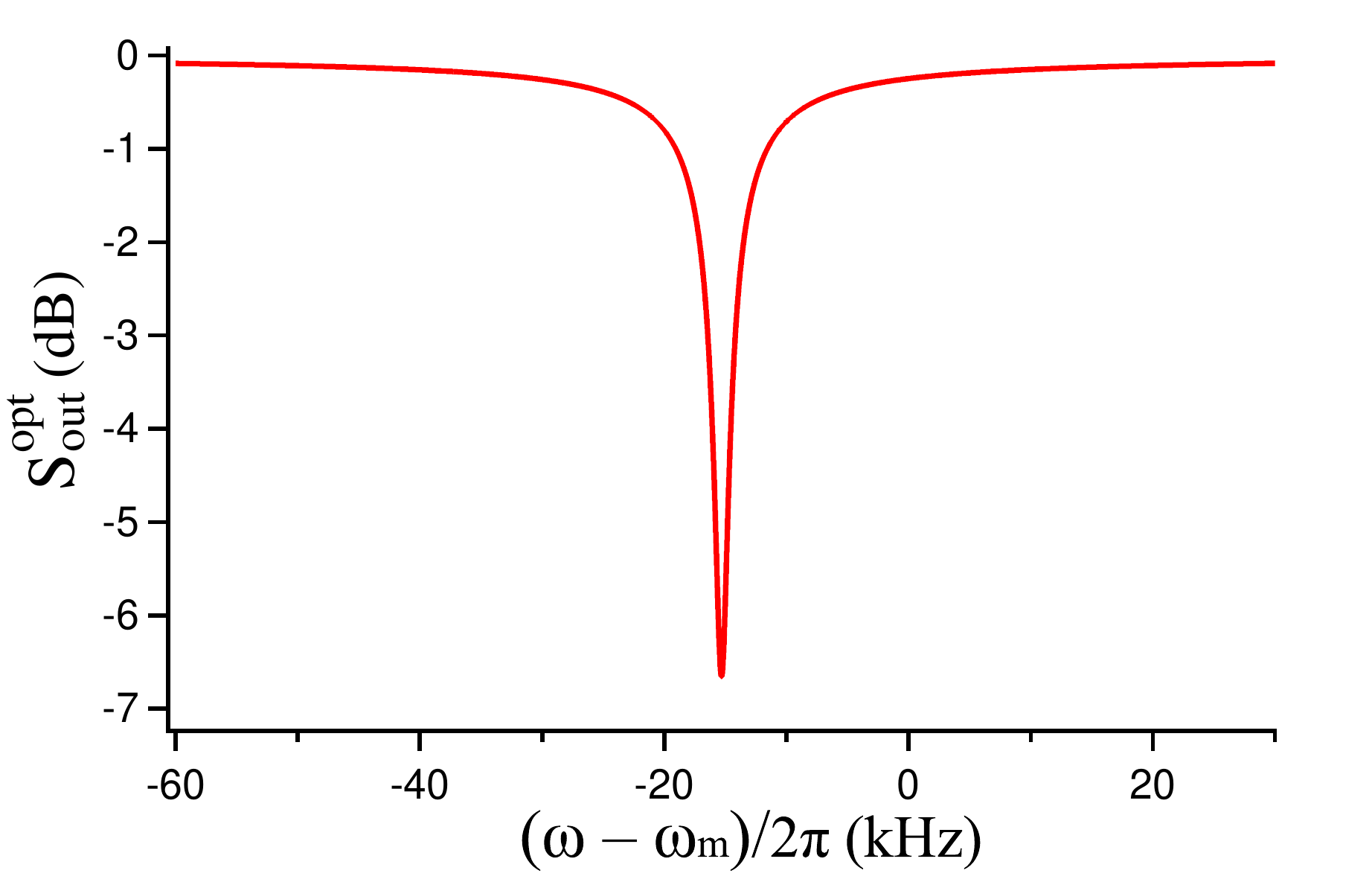}
\caption{Optimal squeezing spectrum of output light in dB $S_{out}^{opt}$. The parameters see text for details.}
\label{Sout}
\end{figure}

Finally, we confirm the quantum fluctuations of optical quadratures outside the cavity. The fluctuations in the output light are determined by the intracavity light and incoming noise. To infer the squeezing and nonclassicality of the light emerging from the cavity, we consider the input-output relations~\cite{noisebook,stable} $b_{out}=\sqrt{2\kappa }b-b_{in}$ (with $b=c, c^\dagger $). In this hybrid atom-optomechanical system atoms contribute to enhancing the quadrature squeezing of output light. The squeezing spectrum can be defined as $S_{out}^{Y}(\omega )=\left( 1/2\pi \right) \int d\Omega e^{-i(\omega +\Omega )t}\left\langle \delta Y_{out}(\omega )\delta Y_{out}(\Omega )\right\rangle $, which is deduced from the frequency-domain correlation function $\left\langle\delta Y_{out}(\omega )\delta Y_{out}(\Omega )\right\rangle=S_{out}^{Y}(\omega )\delta (\omega +\Omega )$. We also note that if $b_{out}$ is in its vacuum, $S_{out}^{Y}(\omega )=1/2\equiv S_{out}^{SN}$ (i.e., the shot-noise limit). However, one usually performs an optimization and considers, for every frequency $\omega $, the optimal squeezing angle possessing the minimum noise spectrum: in this way the optimal squeezing spectrum of output light is defined as~\cite{lightsqueez,hybridreview}
\begin{equation}
\begin{split}
& S_{out}^{opt}\left( \omega \right) = \\
& \frac{2S_{out}^{X}\left( \omega \right) S_{out}^{Y}\left( \omega \right)
-2[S_{out}^{XY}\left( \omega \right) ]^{2}}{S_{out}^{X}\left( \omega \right)
+S_{out}^{Y}\left( \omega \right) +\sqrt{[S_{out}^{X}\left( \omega \right)
-S_{out}^{Y}\left( \omega \right) ]^{2}+4[S_{out}^{XY}\left( \omega \right)
]^{2}}}.
\end{split}
\end{equation}
The optimal values of the parameters are almost the same as those used in Fig.~\ref{intrawigner} {\bf (a)} except the following parameters: effective cavity detuning $\tilde{\Delta}_c\sim2\omega_m$, cavity decay rate $\kappa/2\pi=3\times10^6$ Hz, effective optomechanical coupling $\chi_{e\!f\!f}=10^7$Hz, and a lower temperature $T=2.6$ mK. We consider fewer atoms ($N\sim 10^{2}$) are trapped in the long cavity, which is more implementable in experiment. The optimal squeezing spectrum of output light $S_{out}^{opt}\left( \omega \right) $ is shown in Fig.~\ref{Sout}. The magnitude of squeezing in dB units is given by $10\log _{10}\left(S_{out}^{opt}/S_{out}^{SN}\right) $, and the maximum squeezing is about $6.7$ dB.

For inducing the nonclassicality and enhancing the squeezing, we perform the same excitation counting as used in Fig.~\ref{intrawigner} {\bf (a)}. Similar to the intracavity case, the negative Wigner function is effectively induced and the output squeezing of the light is slightly increased through excitation counting on the atoms. Although the operating temperature is at the order of K, the negativity of the Wigner function is still maintained as the atomic decay rate $\gamma_a$ increases by one order of magnitude. This operating temperature is feasible for experiment, and still much higher than $T\sim10$ mK that can be reached with standard dilution refrigerators. Some robustness is also found against the mechanical decay rate: the Wigner function sustains its negativity against an increase in $\gamma_m$ of about $30\%$. Compared to the condition inside the cavity, the robustness of the nonclassicality with respect to environment disturbances is decreased due to the larger cavity decay.

\section{Conclusions}
\label{concl}

We have studied quantum noise reduction of light in a hybrid atom-optomechanical system focusing on the role of implementing measurements on the atoms. We show that when the system is at steady states with large fluctuations of light much above the SQL, excitation counting on atoms can effectively suppress the noise in one quadrature close to the SQL, but fails to surpass the limit. When the system is at low temperatures, quadrature squeezing of light below the SQL is found with realistic parameters. By increasing the atom-light coupling strength, the squeezing is enhanced with the laser power optimized close to instability. Significant squeezing of the light below the SQL could be achieved by further decreasing various losses in the system. The squeezing below the SQL is benefitted from the presence of atoms and excitation counting on atoms, and can occur at high temperatures. We have also shown that excitation counting on atoms can induce robust negativity of the Wigner function. The results indicate that making non-Gaussian measurements on atoms plays an active role in enhancing optical squeezing and nonclassicality. This study will contribute to the ongoing attempts to prepare non-classical states and also establish atom-light interfaces for continuous variable quantum information processing.

\section{Acknowledgments}
We acknowledge useful discussions with J. Li. This work is supported by National Natural Science Foundation of China (Grants Nos. 61405138, 61505136, 61775158, 61731014, and 11634008), Natural Science Foundation of Shanxi (Grant No. 201701D221116), and Shanxi Scholarship Council of China (Project No. 2017-040). H. Shen acknowledges the financial support from the Royal Society Newton International Fellowship (NF170876) of UK. TCZ acknowledges the support from the National Key Research and Development Program of China (Grant No. 2017YFA0304502).

\renewcommand{\theequation}{A-\arabic{equation}}
\setcounter{equation}{0}
\section*{APPENDIX}
\label{app1}

Here we provide the analytical expression for the characteristic function of the cavity field conditioned on the outcomes of the atomic counting. Having the CM of the system, in virtue of Eqs.~\eqref{66cm} and \eqref{rhosystem} in the main text, the density matrix of the system writes as
\begin{equation}
\rho_{mca}=\frac{1}{\pi^3}\iiint \text{d}^2\alpha\,\text{d}^2\beta\,\text{d}^2\gamma\,\zeta(\alpha,\beta,\gamma)D_m(-\alpha)D_c(-\beta)D_a(-\gamma),
\end{equation}
The density matrix of the conditional state of the cavity field is thus
\begin{equation}
\begin{split}
\rho_c&=n_c{\rm Tr}_{m,a}\left[\,|s\rangle_c\langle s|\,\, \rho_{mca}\right] \\
&=\frac{n_c}{\pi^3}\!\!\iiint\!\!\text{d}^2\alpha\,\text{d}^2\beta\,\text{d}^2\gamma\,\zeta(\alpha,\beta,\gamma){\rm Tr}\left[D_m(-\alpha)\right]D_c(-\beta)\,\langle s|D_a(-\gamma)|s\rangle \\
&=\frac{n_c}{\pi^3}\!\!\iiint\!\!\text{d}^2\alpha\,\text{d}^2\beta\,\text{d}^2\gamma\,\zeta(\alpha,\beta,\gamma)\,\pi\delta^2(\alpha)\,D_c(-\beta)\,e^{-|\gamma|^2/2}L_s^{(0)}(|\gamma|^2) \\
&=\frac{n_c}{\pi^2}\!\int\!\text{d}^2\beta\,D_c(-\beta)\int\!\text{d}^2\gamma\,e^{-|\gamma|^2/2}\,\zeta(0,\beta,\gamma)\,L_s^{(0)}(|\gamma|^2) \\
&=\frac{n_c}{\pi^2}\!\int\!\text{d}^2\beta\,D_c(-\beta)\,G_s(\beta)
\end{split}
\label{rholight}
\end{equation}
where $n_c$ is a constant to guarantee ${\rm Tr}[\rho_c]=1$, and $G_s(\beta)$ is in the form of
\begin{equation}
G_s(\beta){=}\int\!\text{d}^2\gamma\,e^{-|\gamma|^2/2}\,\zeta(0,\beta,\gamma)\,L_s^{(0)}(|\gamma|^2),
\end{equation}
with $L_s^{(l)}(x)$ an associated Laguerre polynomial. Normalization $n_c$ can be determined by ${\rm Tr}[\rho_c]{=}1$, which is $n_c=\pi/G_s(0)$. The characteristic function of the cavity field can then be obtained by
\begin{equation}
\zeta(\lambda)={\rm Tr}[D_c(\lambda)\,\rho_c].
\label{charactlight}
\end{equation}
Substituting Eq.~\eqref{rholight} and $n_c$ into Eq.~\eqref{charactlight}, we finally acquire the characteristic function of light conditioned on the outcomes of measurement $\Pi{=}|s\rangle_a\langle s|$ upon the atomic state
\begin{equation}
\zeta(\lambda)=\frac{G_s(\lambda)}{G_s(0)}.
\end{equation}

\end{document}